\documentclass[graybox]{svmult}

\usepackage{type1cm}        % activate if the above 3 fonts are
\usepackage{makeidx}         % allows index generation
\usepackage{graphicx}        % standard LaTeX graphics tool
\usepackage{multicol}        % used for the two-column index
\usepackage[bottom]{footmisc}% places footnotes at page bottom

\usepackage{newtxtext}       % 
\usepackage{newtxmath}       % selects Times Roman as basic font
\usepackage{url}

\makeindex             % used for the subject index
 \newcommand{\MMM}{\mathcal{M}}
 \newcommand{\NNN}{\mathcal{N}}
 \newcommand{\VF}{\mathfrak{X}}

\newcommand{\YY}{{\boldmath \mbox{$Y$}}}

\newcommand{\uu}{{\boldmath \mbox{$u$}}}

\newcommand{\xx}{{\boldmath \mbox{$x$}}}
\newcommand{\yy}{{\boldmath \mbox{$y$}}}

\newcommand{\XX}{{\boldmath \mbox{$X$}}}

\newcommand{\LL}{{\boldmath \mbox{$L$}}}

\newcommand{\mm}{{\boldmath \mbox{$m$}}}

 \newcommand{\uE}{{^\uparrow}E}
 \newcommand{\dE}{{^\downarrow}E}
 \newcommand{\uS}{{^\uparrow}S}
 \newcommand{\dS}{{^\downarrow}S}
 \newcommand{\dF}{{^\downarrow}F}
 \newcommand{\dL}{{{^\downarrow}L}}
 \newcommand{\dM}{{{^\downarrow}M}}
 \newcommand{\dMM}{{^\downarrow\mathbf{M}}}
 \newcommand{\dLL}{{^\downarrow\mathbf{L}}}
 \newcommand{\uL}{{{^\uparrow}L}}
 \newcommand{\uLL}{{^\uparrow\mathbf{L}}}

\newlength{\defbaselineskip}
\newcommand{\setlinespacing}[1]%
           {\setlength{\baselineskip}{#1 \defbaselineskip}}

\begin{document}

\title*{Learning Physics from Data: a Thermodynamic Interpretation}
\author{Francisco Chinesta, El\'ias Cueto, Miroslav Grmela, Beatriz Moya, Michal Pavelka and Martin {\v S}{\' i}pka}

\authorrunning{Learning physics from data}% for an abbreviated version of
\institute{Francisco Chinesta \at ESI Group Chair. Arts et Metiers ParisTech. 155 Boulevard de l'Hopital. 75013 Paris, France.
\and El\'ias Cueto \at Aragon Institute of Engineering Research. Universidad de Zaragoza. Maria de Luna, s.n. 50018 Zaragoza, Spain.
\and Miroslav Grmela \at \'Ecole Polytechnique de Montr\'eal, C.P. 6079 succ. Centre-ville, Montr\'eal, H3C 3A7 Qu\'ebec, Canada.
\and Beatriz Moya \at Aragon Institute of Engineering Research. Universidad de Zaragoza. Maria de Luna, s.n. 50018 Zaragoza, Spain.
\and Michal Pavelka \at Mathematical Institute, Faculty of Mathematics and Physics, Charles University, Sokolovsk\'a 83, 186 75 Prague, Czech Republic.
\and Martin {\v S}{\' i}pka \at Mathematical Institute, Faculty of Mathematics and Physics, Charles University, Sokolovsk\'a 83, 186 75 Prague, Czech Republic.}

\maketitle

\abstract*{}

\abstract{
	Experimental data bases are typically very large and high dimensional. To learn from them requires to recognize important features (a pattern), often present at scales different to that of the recorded data. Following the experience collected
	in statistical mechanics and thermodynamics, the process of recognizing
	the pattern (the learning process) can be seen as a dissipative time evolution driven by entropy from a detailed level of description to less detailed.
	This is the way thermodynamics enters machine
	learning. On the other hand, reversible (typically Hamiltonian) evolution is propagation within the levels of description, that is also to be recognized. This is how Poisson geometry enters machine learning. Learning to handle free surface liquids and damped rigid body rotation serves as an illustration.
}

\begin{quote}
Helena: Will they be happier when they can feel pain?\\
Dr. Gall: On the contrary. But they will be technically more perfect.
\begin{flushright}
---Karel \v{C}apek, R.U.R. (Rossum's universal Robots).  
\end{flushright}
\end{quote}

\section{Introduction}
An ideal gas that is left undisturbed reaches a state, called an equilibrium state,
at which its behavior is found to be well described by the classical equilibrium
thermodynamics. The features of the ideal gas that play an important role in
the classical equilibrium thermodynamics are thus revealed in the process that
prepares the ideal gas for equilibrium thermodynamics. The equation governing
the time evolution describing the preparation process has been introduced at
the end of nineteen century by Ludwig Boltzmann \cite{Boltzmann}. The equation is now called
the Boltzmann equation. The equilibrium thermodynamics emerges from its
solutions as features of the solutions that survive the dissipation eliminating
gradually in the course of the time evolution the irrelevant details. The ”Natural
Intelligence” (NI), c.f. \cite{Gorban2018_inteligence}, entering the dissipation-driven pattern recognition process is the
realization that the binary collisions are the principal culprits of the disorder
generation that creates the irrelevant details and makes the pattern to emerge.
Due to very fast and very large changes of the gas particle trajectories that occur
during the collisions, details of the trajectories are escaping our attention that is
specified by choosing only the one particle distribution function as the variable
describing states of the ideal gas. Such loss of details enters the Boltzmann
equation as a new dissipative term that breaks the time reversibility of mechanics
and brings solutions eventually to equilibrium states.

Can we see the dissipation-driven pattern recognition process\footnote{In this work we mean \textit{pattern recognition} in a broad sense as a process of extracting any information from the data. The dissipative-driven pattern recognition can be then imagined as a retouche of the original data leading to recognition of the important aspects.} as a result of
a data-driven learning? Let us assume that we have in our disposition trajectories of a large number of gas particles. This is our data base  with which
we begin our investigation. We now apply to the data base  the methods Proper Orthogonal Decomposition (POD), Locally Linear Embedding (LLE),
Topological Data Analysis (TDA), etc. developed in \cite{Kosambi,Golub,LLE,TDA}, for finding a structure in the data base. We
conjecture that such analysis would lead to the same structure as the one revealed in the Boltzmann dissipation-driven pattern recognition process. In other
words, we conjecture that ”Artificial Intelligence” (AI) analysis also reveals the
Boltzmann insight that the binary collisions represent the essential physics involved in the possibility to use equilibrium thermodynamics for describing the
experimentally observed behavior of ideal gases.

We shall use hereafter the following terminology. NI modeling is the "natural intelligence" modeling that has unfolded from Newton's   formulation of mechanics. An NI model  is the time evolution equation (\ref{1}). AI modeling is the ``artificial intelligence'' modeling that is also referred to as machine learning. Data base plays  important but different roles in both NI and AI modeling.

In the NI modeling the data base serves first only as  one of the inspirations leading to a physical insight needed to write down the time evolution equation (\ref{1}) that then represents the NI model. Equation (\ref{1}) is subsequently  solved and its solutions (i.e. predictions of the NI model) are  compared with the data base. The comparison is  the process of the validation of the NI model. The real  beginning of the NI modeling is the time evolution equation (\ref{1}). The NI modeling is rather insight driven than data driven. The data base however participates in the formulation of Eq.(\ref{1}) and
then continues to inspire also the process of solving it (see more in Section 2) and finally it is used to validate the NI model.

In the AI modeling the data base is the principal input. The AI modeling is truly  data driven \cite{Brunton} \cite{kaiser2018discovering} \cite{kevrekidis2010equation}. The objective is to formulate Eq.(\ref{1}) (or an equivalent to it set of instructions for computer that allow to simulate the time evolution) that generates the data base. The beginning of the AI modeling is thus the data base,  its  final result  is the physical insight  introduced  at the  beginning of the NI modeling in the form of Eq.(\ref{1}). In this sense, the AI modeling is a learning process.

In this paper we  recall first (in Section 2) some aspects of the  NI modeling  that, as we show in Section \ref{sec.machine}, play very likely   an important role also in the  machine learning. We focus our attention in particular on the passage from the  time evolution equation  (\ref{1}) to a simple time evolution equation  in which the essential overall features (the pattern in solutions of (\ref{1})) became manifestly displayed and  unimportant details were ignored. Such passage is a principal step in getting an  insight needed to make  predictions based on (\ref{1}). Such passage is also, as we recall in Section 2, a general formulation of thermodynamics. Our principal objective in this paper is thus to contribute to the development of \textit{thermodynamics of machine learning}. In trying to recognize common features in the NI and AI modeling we follow the spirit of  Machine Learning via Dynamical Systems
proposed in \cite{E,E2018}.

The analysis presented in Sections 2 and 3 are illustrated in Section \ref{sec.ill} on the example of free surface fluid flows discussed already in \cite{sloshing} and in Section \ref{sec.RB} on learning from evolution of a freely rotating damped rigid body. It is shown that by learning only mechanics (rigid body without dissipation), one can not learn the enrgy completely. The problem is that the energy can be shifted by a Casimir invariant (here magnitude of the angular momentum) leaving the mechanics unaltered. However, when learning also dissipation, the Casimirs are learned as well, and one can infer the whole expression for energy.

Novelty of this paper lies in the following points.
 Learning is dictated by entropy production, i.e. removing details and capturing first order insights. This is the main aim of dimensionality reduction (linear and nonlinear). When learning physics, thermodynamics is the appropriate framework for accomplishing it safely and precisely. This provides a thermodynamic interpretation of the rather numerical approach from \cite{sloshing}.
It is moreover important to recognize both the projection and the inverse embedding between the different detailed and less detailed manifolds (scales), as within the MaxEnt framework. We learn from detailed data, by removing details, etc. Then we predict in the reduced space, in which we created our (reduced) model, but we validate in the rich space, and for that the embedding is needed. At least when addressing physics, both scales are thermodynamically linked and we move from one to the other for coming back later. This thermodynamic link can be exploited in the numerical algorithms. Moreover, learning can be enhanced by recognition of the geometric structure generating the evolution, as for instance in Sec. \ref{sec.RB} when learning kinetic energy of rigid body from trajectory of its angular momentum.

\section{Pattern recognition in statistical physics and thermodynamics}\label{sec.reduction}

In this section we recall some ideas and methods that have emerged in statistical mechanics and thermodynamics and that, as shown in Section \ref{sec.machine}, are also pertinent in machine learning.

\subsection{Reduction and pattern recognition}

Consider a manifold $\MMM$ with coordinates $x\in \MMM$, and assume that there is a vector field $\XX\in \VF(\MMM)$ on the manifold. The vector field determines a flow on the manifold. In other words, components of the vector field are the right hand sides of evolution equations for $x$,
\begin{equation}\label{1}
	\dot{x}^i = X^i(x),
\end{equation}
and the evolution simply follows arrows of the vector field. We note that if the system under investigation is a physical system composed of atoms and molecules, then one possible model in the form of Eq.(\ref{1}) is in principle known. The state variable $x$ consists of the position vectors and momenta of all the particles involved (provided we limit ourselves to the classical mechanics) and the vector field $\XX$ is the vector field of classical mechanics (right hand side of Hamilton canonical equations). To specify it we need to know (or assume to know) all the forces participating  in the time evolution.  If the system under investigation is still a physical system, but the data base addresses some macroscopic features (e.g. fluid flows), then $x$ has to address the quantities entering the data base and an additional insight is needed to formulate the vector field $\XX$.

Now we turn to the problem of solving  Eq. (\ref{1}), i.e. to the problem of finding the flow generated by (\ref{1}).  There are two routes to follow. On the first route we find all details of the trajectories generated by (\ref{1}). This, of course, is in general a very difficult task even for very well performing computers. Moreover, the result,  i.e. the phase portrait generated by (\ref{1}), still needs to be subjected to a pattern recognition process in order to be useful. The complexity of the phase portrait has to be reduced by highlighting  important features and ignoring  unimportant details. On the second route the objective is not to find all the details of  solutions of Eq. (\ref{1}) but only their important qualitative features.  We shall follow the second route.

Consider  a projection $\pi:\MMM\rightarrow\NNN$, range of which determines a reduced manifold $\NNN$. An insight (inspired also by the data base in our disposition) is needed to specify the projection $\pi$. As an example, we take (\ref{1}) to be the Boltzmann kinetic equation (i.e. $x$ is the one particle distribution function) and $\pi$ the projection to hydrodynamic fields (that are the first five moments of the distribution function in the velocity variable).

The projection $\pi$ maps each point $x\in\MMM$ to a point $y\in\NNN$. To each point $x$ there is an arrow attached (vector field $\XX$), and this arrow (an instruction how to proceed in the time evolution in $\MMM$) can be also mapped to the tangent bundle of $\NNN$, i.e. to a vector tangent to $\NNN$ attached to a point $y\in\NNN$. The projected vector field then generates the time evolution in $\NNN$.
However, in a thermodynamic setting---this is not the case in projection-based model reduction---there are typically many points from $\MMM$ projected to single $y\in\NNN$, there are many vectors to be attached to $y$. How to choose the right one (i.e. the one  expressing properly the induced flow on $\NNN$) and consequently how to determine the vector field on $\YY\in\VF(\NNN)$,
\begin{equation}
	\dot{y}^a = Y^a(y)
\end{equation}
representing the \textit{reduced dynamics}?

In order to  answer this question we need again an insight.
Imagine  a phase portrait where trajectories of a dynamical system are depicted. For a physical system it is usually possible to find a pattern where typical trajectories are contained, see e.g. \cite{Netocny2002}. When starting somewhere in the phase space, the point typically evolves towards the pattern. The  reduction introduced above  takes the phase space (or the vector field generating it) and finds a reduced manifold where typical evolution takes place, i.e. leads to the pattern recognition. In particular, geometry of the reversible evolution on $\NNN$ is inherited from the geometry on $\MMM$. It is therefore not surprising to anticipate (see more in Section \ref{sec.machine}) that dynamic reductions provide inspiration for machine learning and vice versa. Let us now recall several methods of the dynamic reduction.

\subsection{Reducing dynamics, thermodynamics}\label{sec.GENERIC}

We begin with an example. Let $\MMM$ be the state space of kinetic theory (i.e., the physical system under investigation is a gas and $x$ is the one-particle distribution function) and $\NNN$ is the state space of the classical equilibrium thermodynamics (i.e., $y=(V,N,E)\in \mathbb{R}^3$, where $V$ is the volume of the region in $\mathbb{R}^3$ in which the gas under investigation is confined, $N$ is the number of moles, and $E$ is the total energy of the microscopic particles composing the gas). In this case, no time evolution takes place in $\NNN$. The projection $\pi$ is thus the projection on the fixed points of the time evolution taking place in $\MMM$.
Let us  assume that the models in $\MMM$ and in $\NNN$ have been validated by their corresponding data bases.

The question that we ask now is of  what we have learned by relating the two models, i.e., by reducing the model in $\MMM$ to the model in $\NNN$. If the model in $\NNN$ was  a  model with the time evolution then we would clearly obtain the time evolution in $\NNN$ as a reduced dynamics and thus learn how to see the time evolution in $\NNN$ from the point of view of $\MMM$. But in the case when the model in $\NNN$ is the equilibrium thermodynamics (i.e.,  there is no time evolution  in $\NNN$,  there is no reduced dynamics)  the question becomes    particularly  pertinent.   Following Boltzmann, we answer the question as follows. A part of the data base corresponding to the kinetic theory is an observation of the  process that prepares the gas under investigation to states at which its behavior can be well described by the model in $\NNN$. According to Boltzmann, the time evolution describing the preparation process is governed by the Boltzmann equation. It is the time evolution generated by the Boltzmann equation that makes the projection $\pi$. We call the dynamics making the projection $\pi$ a \textit{reducing dynamics}. The dynamics expressed in the Boltzmann equation is thus an example of reducing dynamics. Following solutions to the Boltzmann equation,  kinetic theory becomes reduced to  equilibrium thermodynamics.

The potential driving the reduction  is called an entropy in $\MMM$. We shall call it an \textit{upper entropy} $\uS$.  This potential,  if evaluated at the  states in $\MMM$ reached as $t\rightarrow \infty$, becomes the entropy in $\NNN$, called a \textit{lower entropy} $\dS$. In the case of $\NNN$ being the state space of the equilibrium thermodynamics,  $\dS$ is  the entropy $S(V,N,E)$ entering the model in $\NNN$. The reduction from $\MMM$ thus gives us the fundamental thermodynamic relation in $\NNN$.

Following \cite{go,og,hco,pkg}, the reducing dynamics to the equilibrium thermodynamics is expressed mathematically by the General Equation for Non-Equilibrium Reversible-Irreversible Coupling, GENERIC,
\begin{equation}\label{eq.Gen}
	\dot{x}^i = \uL^{ij} \frac{\partial \uE}{\partial x^j} + \frac{\partial \Xi}{\partial x^*_i}\Big|_{x^*_i = \frac{\partial \uS}{\partial x^i}}.
\end{equation}
The Boltzmann kinetic equation as well as many other equations (e.g., the Navier-Stokes-Fourier of fluid mechanics) expressing dynamics in other state spaces $\MMM$ (see  \cite{go,og,hco,pkg}) are particular examples of Eq. (\ref{eq.Gen}). We now explain the meaning of the symbols appearing on the right hand side of Eq. (\ref{eq.Gen}).

The first part of the right hand side is the Hamiltonian evolution, constructed from the Poisson\footnote{The Poisson bracket corresponding to the Poisson bivector is $\{F,G\}=\langle F_x| \uLL| G_x \rangle$, where $\langle\bullet|\bullet\rangle$ denotes a scalar product.} bivector $\uLL$ and the gradient of energy $\uE$. Hamiltonian dynamics conserves energy (due to the antisymmetry of $\uLL$) and entropy (due to the requirement that $\uS$ is the Casimir of Poisson bracket, i.e., the requirement that
\begin{equation}
	\uL^{ij}\frac{\partial \uS}{\partial x^j} = 0 \qquad \forall i.
\end{equation}

The second term in Eq.(\ref{eq.Gen}) is a gradient dynamics, where $x^*_i$ are conjugate variables and $\Xi(x,x^*)$ is a dissipation potential with convex dependence on them  (see more in \cite{pkg}). From the convexity it follows that
\begin{equation}
	\dot{\uS} = \left(x^*_i  \frac{\partial \Xi}{\partial x^*_i}\right)\Big|_{x^*_i = \frac{\partial \uS}{\partial x^i}}\geq 0,
\end{equation}
Moreover, the dissipation potential $\Xi$ and entropy $\uS$ have to be such that energy $\uE$ is conserved in the gradient dynamics. These properties of $\uE,\uS,\uLL,\Xi$, together with the convexity of $\uS$ and the requirement that $\Xi$ reaches its minimum at  $x^*=0$,  makes it possible to regard $(-\uS)$ as a Lyapunov function displaying the approach, as $t\rightarrow \infty$, to the equilibrium states at which the entropy $\uS$ reaches its maximum. Such states then form $\NNN\subset\MMM$ (in the sense that $\NNN$ be isomorphic to a submanifold of $\MMM$).
The Hamiltonian mechanics is moreover reversible with respect to time-reversal transformation while gradient dynamics is irreversible \cite{PRE15}, and generalized Onsager reciprocal relations \cite{Onsager1930,Onsager1931,Casimir1945,dGM} are automatically fulfilled, see \cite{hco,pkg,RedExt}.

Let us assume now that we are projecting from $\MMM$ to $\NNN$ on which the time evolution  still takes place. In the next subsection we shall discuss the reduced dynamics, i.e. the projection of the vector  field $\XX  \in \VF$  on the the vector field $\YY\in \VF(\NNN)$. For a moment, we assume that the reduced dynamics is known.
It has been conjectured in \cite{MG-CR,Miroslav-guide} that (\ref{eq.Gen}) with an appropriate modifications of the properties required from  $\uE,\uS,\uLL,\Xi$,   expresses also reducing dynamics to $\NNN$ on which the time evolution takes place. In such case, the result of the dynamic reduction is the \textit{reduced dynamics} (that we discuss in more detail in the next subsection) and thermodynamics in $\NNN$ that is inherited from the entropy $\uS$ generating the reducing time evolution leading from $\MMM$ to $\NNN$.

Summing up, we see that the dynamical reduction from $\MMM$ to $\NNN$, that can be seen as a process of learning the model in $\NNN$ from the model in $\MMM$, makes possible to see the dynamics in $\NNN$ as a reduced dynamics from $\MMM$ and, in addition, introduces into $\NNN$ a new element that has been absent in the original model in $\NNN$. The new element is thermodynamics. It is the fundamental thermodynamic relation in $\NNN$ expressed in the entropy $\dS$. If the model in $\NNN$ is the equilibrium thermodynamics, then the fundamental thermodynamic relation arising in the dynamical reduction is the fundamental thermodynamic relation constituting  the model in $\NNN$ (i.e. the equilibrium thermodynamics). If, on the other hand, the model in $\NNN$ involves the time evolution, then such model does not (at least in general) involve any thermodynamic relation and thus the fundamental thermodynamic relation arising in the dynamic reduction is a new information obtained  from seeing the model in $\NNN$ from the point of view of the more detailed model in $\MMM$.  

Still another thermodynamics  in $\NNN$ arises  if we regard the model in $\NNN$ as a more detailed than another model in $\verb"N"$. The upper entropy $\uS$ appearing in (\ref{eq.Gen}) with $x$ replaced by $y$, i.e. the upper entropy $\uS$ generating the time evolution from $\NNN$ to $\verb"N"$, introduces thermodynamics in $\NNN$ (that is different from the thermodynamics introduced by $\dS$) obtained from seeing the model in $\NNN$ as a basis for reduction to a less detailed model in $\verb"N"$.

Finally, we note that if we are interested only in the result of the time evolution generated by (\ref{eq.Gen}), then we can replace (\ref{eq.Gen}) by simply a MaxEnt reduction which consists of the maximization of the upper entropy $\uS$ subjected to the constraints $\pi (x)$, as shown in the appendix of \cite{RedExt}. The Lagrange multipliers in this maximization are $y^*$. This is indeed the principle of maximum entropy (MaxEnt) formulated by Shannon \cite{Shannon} and Jaynes \cite{Jaynes}. The question that arises in this static viewpoint of the reduction is of what is the entropy $\uS$, how shall we find it. In the dynamical viewpoint the upper entropy $\uS$ is the potential generating the reducing time evolution (that is, in general, a part of the data base associated with the model in $\MMM$). In the static viewpoint of the reduction one has to turn to other insights (see \cite{Shannon} and \cite{Jaynes} for more details).

\subsection{Reduced dynamics}\label{sec.Turkington}

We turn our attention now to the reduced dynamics, i.e., to the projection of $\XX\in\VF$ to $\YY\in\VF(\NNN)$.

Perhaps the simplest method of projecting $\XX\in\VF$ to  $\YY\in \VF(\NNN)$ is provided by MaxEnt.  Pick one point $y\in\NNN$. Due  to the MaxEnt embedding there is an associated point $\pi^*(y)\in\MMM$. Take the vector attached to that point and project it to $y$. The vector field $\YY\in\VF(\NNN)$ obtained by repeating this for each $y\in\NNN$ is the MaxEnt projection of $\XX$ onto $\NNN$
\begin{equation}
	Y^a(y) = \frac{\partial \pi^a}{\partial x^i}\Big|_{x(y)}X^i(x(y)).
\end{equation}
But this vector field has a drawback. The trajectories obtained by solving evolution equations $\dot{y} = \YY$  approximate poorly the trajectories  on the $\MMM$ manifold. This is because the approach towards states with higher entropy is not explicitly contained in $\YY$. Therefore, a more precise approximation is needed, see \cite{Gorban}, \cite{DynMaxEnt}.

A classical example of reduction beyond MaxEnt is the Chapman-Enskog expansion \cite{Chapman-Cowling}. Let $\MMM$ be the state space of kinetic theory (i.e., the physical system under investigation is a gas and $x$ is the one-particle distribution function) and $\NNN$ is the state space of the hydrodynamics (i.e., hydrodynamic fields of density, momentum density and energy density, $y=(\rho,\uu,e)$). 
In this case, the time evolution takes place in $\NNN$ is often well described by the Navier-Stokes-Fourier system of equations, see e.g. \cite{dGM}, obtained by the Chapman-Enskog expansion. The projection $\pi$ is the projection on the first 5 moments of the distribution function, and the detailed Boltzmann equation (vector field $\XX$) is reduced to less detailed Navier-Stokes-Fourier equations (vector field $\YY$). The upper entropy $\uS$ is the Boltzmann entropy and it generates a lower-level entropy $\dS$, expressed by the Sackur-Tetrode relation for ideal gases \cite{callen,pkg}. 
The embedding $\pi^*$ is the MaxEnt mapping from hydrodynamic fields to the locally Maxwellian distribution functions. The locally Maxwellian distribution functions form the local equilibrium submanifold of $\MMM$, which is isomorphic to $\NNN$. When the evolution in $\MMM$ takes place close to the local equilibrium submanifold, the evolution in $\NNN$ is close to the detailed evolution in $\MMM$. The Chapman-Enskog expansion, however, also has a few drawbacks. Firstly, it relies on the a priori unknown form of asymptotic expansion and, secondly, it requires the presence of dissipative terms in vector field $\XX$.

Another robust method of projecting the vector field $\XX$ to $\YY$  was formulated by Bruce Turkington in \cite{Turkington}. The reduction consists of the following steps. Consider a manifold $\MMM$. Liouville equation for the probability distribution function on the manifold is formulated, and linear projection from the distribution function is defined, range of which determines a manifold $\NNN$. Shannon entropy is assumed for the distribution function, which forms and embedding $\pi^*$ of $\NNN$ onto $\MMM$.

Let us first project Hamiltonian mechanics on $\MMM$ (the Liouville equation) to Hamiltonian mechanics on $\NNN$. The upper\footnote{The more detailed level is referred to  as the upper while the less detailed (reduced) as lower.} Poisson bivector $\uLL$ is projected as a twice contravariant tensor field on the space of state variables and, if necessary, evaluated at the MaxEnt embedding,
\begin{equation}
	\dL^{ab} = \left(\frac{\partial \pi^a}{\partial x^i} \uL^{ij}(x) \frac{\partial \pi^a}{\partial x^i}\right)\Big|_{x=\pi^*(y)}.
\end{equation}
To construct the Hamiltonian vector field on the lower level one further needs a Hamiltonian, energy on the lower level.

Let energy on $\MMM$ be $\uE(x)$. Energy on the lower level $\NNN$ is inherited from the higher level through the MaxEnt mapping $\dE(y) = \uE(\pi^*(y))$. However, since some energy modes present on the higher level have already been damped on the lower level, typically $\dE(\pi(x))\neq \uE(x)$. If the latter relation were an equality, one could project the higher-level evolution to the lower-level easily as one would obtain that time derivative of $\pi(x)$ be equal to $\dLL \cdot d\dE$, which would be the lower-level purely Hamiltonian vector field. Since, however, the equality typically does not hold, simple projection does not give the desired result.

Instead, a lack-of-fit Lagrangian is defined which compares projections of the exact trajectories on $\MMM$ with trajectories on $\NNN$. Minimization of the Lagrangian then leads to a GENERIC evolution on $\NNN$ and gives a dissipation potential driving thermodynamic evolution on $\NNN$. The method has recently been generalized in \cite{JSP2020}.

Still another method of constructing the reduced vector field is
the Ehrenfest method developed in \cite{Ehrenfests,GK-Ehrenfest, GK-Ehrenfest2} and \cite{Gorban}. The method  has the following ingredients: detailed manifold $\MMM$ equipped with entropy and with a vector field (evolution equations), manifold $\NNN$ and projection $\pi$ from $\MMM$ to $\NNN$. MaxEnt then provides the embedding of $\NNN$ into $\MMM$ as usually. The vector field on $\MMM$ does not need to have the GENERIC structure, but it is advantageous as shown in \cite{Pavelka-LD}.

The vector field $\XX\in\VF(\MMM)$ is first projected to a vector field $\YY_0\in\VF(\NNN)$ by the MaxEnt projection. This vector field, however, needs to be corrected. Therefore, the vector field $\XX$ is lifted to the tangent bundle $T\MMM$ and subsequently projected back to $\MMM$, which results in a smoothed vector field on $\MMM$, $ER(\XX(\MMM))$, which expresses a sort of overall motion on $\MMM$, called Ehrenfest regularization in \cite{ehrenfest-regularization}. The same is done with vector field $\YY_0$, which results in vector field $ER(\YY_0)\in\VF(\NNN)$. Finally, vector field $ER(\XX)$ is MaxEnt-projected to $\NNN$ and compared with $ER(\YY_0)$. A correction term is then added to $\YY_0$, forming a new vector field $\YY_1\in\VF(\NNN)$, which makes $ER(\XX)$ equal to $ER(\YY_1)$ (to a given order of relaxation time parameter). Vector field $\YY_1$ then represents the evolution on $\NNN$, its components are right hand sides of evolution equations for $y\in\NNN$. This is the Ehrenfest reduction of detailed evolution on $\MMM$.

Another method of dynamic reduction is the Dynamic MaxEnt developed in \cite{Grmela2013-CMAT,RedExt,DynMaxEnt}. The main idea is to first promote the conjugate variables $x^*$ in the GENERIC framework (\ref{eq.Gen}) to independent variables, which is natural from the point of view of contact geometry \cite{Miroslav2014-Entropy,pkg}. The goal is to reduce a GENERIC model for state variables on manifold $\MMM$ so that a fast variables relaxes and becomes enslaved by the remaining slower variables, $\NNN$ being the manifold of slow variables.

The fast variable is first evaluated at the MaxEnt value determined by the remaining state variables. But since the conjugate fast variable is still present in the evolution equations for the slow variables, we need to express the conjugate variable in terms of the remaining state and conjugate variables. The fast conjugate variable is found as the solution to the evolution equation of the fast state variable evaluated at the MaxEnt value of the state variable. The conjugate fast variable is thus determined by compatibility of the MaxEnt value of the fast variable and the evolution equation for the fast variable. This way we end up with a vector field for the slow variables (on manifold $\NNN$) compatible with the MaxEnt embedding of the slow manifold into the original manifold.

\section{Pattern recognition in machine learning}\label{sec.machine}

Imagine now a robot \cite{RUR} that is, for instance, supposed to perform a mechanical task with a physical system, as e.g. in \cite{sloshing}, while learning by itself how to operate the system.
The robot has as the input a set of discrete trajectories on $\MMM$, $G(\MMM)$. It should give as output an approximation of them by a low dimensional vector field which can be used to predict future evolution of the system approximately (so that it can be operated in a reasonable way).

\subsection{General scheme}

For simplicity we shall illustrate the machine learning using the Proper Orthogonal Decomposition (POD), but the general picture will be applicable also to other methods. The problem is that the robot has discrete trajectories on a high-dimensional manifold $\MMM$, and it would be too costly to reconstruct the vector field $\XX\in\VF(\MMM)$ from them; the vector field would have too many dimensions. Moreover, such high dimensional model would not provide the insight we look for.
The trajectories must  be approximated by trajectories on a low dimensional manifold $\NNN$. Therefore, the task consists of the following steps:
\begin{enumerate}
	\item \textit{Manifold recognition}: Find a low-dimensional manifold $\NNN$ such that a projection of trajectories $G(\MMM)$ to $\NNN$ well approximates the original set $G(\MMM)$ of trajectories on $\MMM$. To accomplish this task, the robot needs the following:
	\begin{enumerate}
		\item To measure distances and define orthogonality, the robot needs a \textit{metric} on $\MMM$, $g(\bullet,\bullet)$, e.g. the $l^2$ scalar product with or without weights.
		\item Find a \textit{projection} operator $\pi:\MMM\rightarrow \NNN$.
		\item To compare in $\MMM$ the trajectories $G(\MMM)$ with their projections to $\NNN$, which is the means of assessing ``goodness'' of the approximative manifold $\NNN$, the robot needs an \textit{embedding} mapping $\pi^*:\NNN\times\dots\times\NNN\rightarrow \MMM\times\dots\times\MMM$, mapping trajectories on $\NNN$ to trajectories on $\MMM$. The embedding is typically determined by MaxEnt in thermodynamics, but it is often difficult to construct it outside thermodynamics. Alternatively, the robot can compare the trajectories on the reduced manifold $\NNN$, for which the embedding is not needed. On the other hand, the embedding will be needed in the last step below anyway.
	\end{enumerate}
	\item \textit{Recognition of the reduced vector field}: Once having the low-dimensional manifold and projected trajectories $\pi(G(\MMM))$, the goal is to find a vector field $\YY \in \VF(\NNN)$ approximating the trajectories on $\NNN$. This is done by choosing an Ansatz on the form of the vector field, e.g. GENERIC, and fitting the unknown parameter so that the trajectories on $\MMM$ and $\NNN$ coincide in a sense. Once this step is successfully finished, the robot has recognized how the typical trajectories are created, he has learned how the system works.
	\item To use this acquired knowledge, the robot is then supposed to integrate the vector field $\YY$ to future times in order to \textit{predict} future states on the $\NNN$ manifold. These states are then embedded into the $\MMM$ manifold of experimental data by mapping $\pi^*$ to obtain prediction of future states of on manifold $\MMM$.
\end{enumerate}
Note that steps 1 and 2 can be seen as pattern recognition (manifold recognition and vector field recognition).

\subsection{Reduced manifold recognition by POD}

Let us now demonstrate the first step (manifold recognition) on a standard reduction method---the proper orthogonal decomposition (POD) or principal component analysis (PCA), see e.g. \cite{Chatterjee,Golub}.

\subsubsection{Loss of information}

Let us have $N$ time snapshots of $m$-dimensional experimental data, assuming that $m>>N$, ordered to a $N \times m$ matrix $Z$. This matrix represents the high dimensional trajectories on manifold of the data $\MMM$. This matrix is now to be approximated by POD. The core of POD is the singular value decomposition of matrix $Z$,
\begin{equation}\label{eq.SVD}
	 Z = U \Sigma V^T,
\end{equation}
where $U$ is an orthogonal $N\times N$ matrix, $V$ is an $m\times m$ orthogonal matrix and $\Sigma$ is an $N\times m$ matrix with entries only on the diagonal. The entries are the called singular values, they are non-negative and ordered, $\sigma_1\geq \sigma_2\geq \dots \sigma_N$. Note that there  no information has been lost so far. The singular values are calculated as square roots of eigenvalues of the symmetric positive definite $N\times N$ matrix
\begin{equation}
	Q = Z Z^T = U \Sigma \Sigma^T U^T.
\end{equation}
In this way we also obtain the matrix $U$,  which consists of the eigenvectors of $Q$.
Now only $k$ first singular values are taken into account while setting $\sigma_l=0$ for all $l>k$, which turns $\Sigma$ to a new matrix $\bar{\Sigma}$. This is the crucial point where reduction takes place. The advantage of SVD is that it gives the best possible $k$-dimensional approximation of $Z$ provided the $l^2$ metric is used.

\subsubsection{Projection}
Finally---look at Eq. (\ref{eq.SVD})---, the relevant part (first $k$ rows, since other are multiplied by zeroes) of matrix $V$ is calculated from $U^T Z = \bar{\Sigma} V^T \stackrel{def}{=}B$. There are $k$ non-zero rows of this $N\times m$ matrix $B$, and these rows, denoted as $v_j\in \mathbb{R}^m$, form a basis of the $k$-dimensional submanifold $\NNN\subset\MMM$. Step (a) in the above abstract procedure is given made by choosing the usual $l^2$ scalar product and corresponding Frobenius norm. Step (b) is made by orthogonal projection $\pi$ to the basis of $\NNN$,
\begin{equation}
	\NNN\ni y = \sum_{j=1}^k \langle x, v_j\rangle v_j \quad\forall x\in \MMM.
\end{equation}

\subsubsection{Embedding}
Consider now a trajectory $(y_1,\dots,y_N)$ on $\NNN$. We construct an $N \times m$ matrix $Y(y_1,\dots,y_N)$ rows of which correspond to $y_i = \sum_{j=1}^k c^j_i v_j$. The embedding $\pi^*$ is then given by
\begin{equation}
	\pi^*(y) = U Y(y),
\end{equation}
which is a trajectory on $\MMM$. Step (c) has been finished. The POD method took the set of trajectories on $\MMM$, encoded it into matrix $Z$, and identified a $k-$dimensional submanifold $\NNN\subset\MMM$ that  approximates  the trajectories $Z$. Moreover, there is an orthogonal projector $\pi$ onto the basis of $\NNN$ and an embedding $\pi^*$ mapping trajectories on $\NNN$ to trajectories on $\MMM$. Step 1, manifold recognition, is thus finished.

\subsubsection{Thermodynamics}\label{sec.eigen.thermo}

We shall now look at  the reduction described above through the eyes of thermodynamics recalled  in
Sec. \ref{sec.GENERIC}.  We regard the embedding of
the  projected manifold $\NNN$ to the original $\MMM$ as a result of a \textit{learning time evolution} which has revealed the important features in the data base collected in $\MMM$. We thus interpret the reducing time evolution  as the learning time evolution. This   dissipative  evolution is generated by an entropy. Having the entropy and focusing our interest only on the final outcome of the learning time evolution,  we can also see the passage from $\MMM$ to $\NNN$ as  maximization of the entropy (MaxEnt principle). We now proceed to identify the entropy associated with POD.

The crucial step in POD where information is lost is the dropping of eigenvalues. We shall seek its thermodynamic interpretation.
A way to calculate eigenvalues is based on minimization of the Rayleigh quotient in a dynamical system\footnote{Another dynamical system converging to eigenvalues of a matrix was found in \cite{Brockett1988}, where the double bracket dissipation, geometrized in \cite{Bloch-dissipation}, was found.}, see \cite{Absil}. Let us interpret the Rayleigh quotient as entropy of a vector related to a matrix $A$,
\begin{equation}\label{eq.eigen.S}
	S(x) = \frac{ x^T\cdot A\cdot x}{x^T\cdot x}.
\end{equation}
Gradient dynamics of $x$ is then prescribed as
\begin{equation}\label{eq.eigen.dyn}
	\dot{x} = \frac{\partial \Xi}{\partial x^*}\Big|_{x^* = S_x}
	= \tau  \frac{\partial}{\partial x} \frac{ x^T\cdot A\cdot x}{x^T\cdot x},
\end{equation}
for $\Xi = \frac{1}{2}\tau (x^*)^2$. The magnitude of $x$ is conserved by the dynamical system, so we can regard $x$ to be normalized to unity. This dynamical system has stationary points corresponding to eigenvectors of matrix $A$, and as it converges to the stationary values, it converges to the eigenvectors. From the eigenvectors, the eigenvalues can be recovered as the Rayleigh quotients, i.e. as the values of entropy in the stationary states. Eigendecomposition can be seen as result of a thermodynamic evolution.

However, as the matrix has typically more eigenvectors, the dynamical system (\ref{eq.eigen.dyn}) has more stationary points. Typically it converges to the eigenvector corresponding to the dominant eigenvalue (highest entropy), but there are other lower eigenvalues (lower entropy) that also represent stationary solutions of the system. By being restricted only to some region around the global maximum of entropy, we obtain the information loss from POD.

Finally, the projection from all vectors normalized to unity (manifold $\MMM$) to the chosen eigenvectors (manifold $\NNN$, also represented by the eigenvalues) is simply the usual orthogonal projection to the span of the eigenvectors. Since the eigenvectors are contained in the original manifold $\MMM$, the embedding is trivial (identity).

The reduction by POD, where only part of spectrum is considered while the remaining eigenspaces being ignored, can be seen as a dynamic reduction driven by entropy and implying a maximum entropy principle.

\subsubsection{Comparison with Locally Linear Embedding}
Locally linear embedding (LLE) \cite{LLE} typically provides better approximation of the low-dimensional manifold than POD. Let us therefore briefly mention the method. Starting with points $x\in\MMM$, a weight matrix $W_{ij}$ is found, which provides local interpolation of points on $\MMM$ by their chosen number of neighbors. Then points $y\in\NNN$ are found as the points that are best interpolated by weights $W_{ij}$. This provides the projection $\pi:\MMM\rightarrow\NNN$. 

How to construct the embedding $\pi^*:\NNN\rightarrow \MMM$? We see three possible routes: (i) One can use a crude interpolation between $y$ and $x$, as e.g. in \cite{sloshing}. (ii) One can reverse the LLE procedure. Starting with points on $\NNN$, constructing new weights $\bar{W}_{ij}$ and finding $x\in\MMM$ that are best interpolated by the new weights, as suggested in \cite{LLE}. (iii) Finally, one can reformulate the LLE projection as gradient dynamics maximizing an entropy. The embedding could be then constructed by the MaxEnt procedure with respect to that entropy. Let us comment on this possibility in more detail.

The LLE algorithm consists of two steps, namely finding the weights $W_{ij}$ and subsequently finding the projection $\pi$. Both the steps are formulated as minimizations of certain cost functions. It can be therefore anticipated that LLE can be reformulated as gradient dynamics. The first step stands for minimization of cost functions
\begin{equation}
	\epsilon_i(W) = (x_i - \sum_j W_{ij}\eta_j)^2,
\end{equation}
where $x_i$ is the $i$-th vector from $\MMM$ and $\eta_j$ is the $j$-th, $j=1,\dots,K$, neighbor of $x_i$. Note that the choice of $K$ and the notion of distance (metric on $\MMM$) are needed. Moreover, the weights are supposed to sum to one for each $i$, $\sum_j W_{ij}=1$, since this is the gauge freedom of the cost function. By minimization subject to the sum-to-one constraint one obtains 
\begin{equation}\label{eq.Wij}
	W_{ij} = -\lambda_i \sum_k C^{-1 (i)}_{jk} + \sum_l x_i \cdot \eta_l C^{-1 (i)}_{jl}
\end{equation}
with $C^{(i)}_{jk} = \eta_j \cdot \eta_k$ being the correlation matrix, $C^{-1 (i)}$ is its inverse, and $\lambda_i = \frac{x_i\cdot \sum_j \eta_j \cdot \sum_k C^{-1 (i)}_{kj}}{\sum_{jk}C^{-1 (i)}_{jk}}$ being the Lagrange multiplier.

The second step is minimization of cost function 
\begin{equation}
	\phi(y) = \sum_i (y_i - \sum_j W_{ij} y_j)^2
\end{equation}
subject to the constraints that $\sum_i y_i =0$ and $y_i \otimes y_j \propto \mathbf{I}$, $\mathbf{I}$ being the $d\times d$ identity matrix on the low-dimensional manifold. This step can be seen as eigenvalue decomposition, and $d$ eigenvectors are then the sought vectors $y_i$, see \cite{LLE} for more details.

Therefore the LLE projection can be seen as eigendecomposition of matrix $W_{ij}$ given by equation \eqref{eq.Wij}. 
It has already been noted in Sec. \ref{sec.eigen.thermo} that eigendecomposition can be seen as gradient dynamics, which means that LLE itself can be seen as gradient dynamics with entropy \eqref{eq.eigen.S} for matrix \eqref{eq.Wij} subject to the constraints imposed on $y$. 

The LLE projection can be seen as gradient dynamics with its own entropy. Let us now assume that a position on the low-dimensional manifold $y\in \NNN$ is known. Can the entropy lead to a consistent construction of the embedding $\pi^*$? We do not know the answer, but we would like to attract attention to this question.

\subsection{Reduced vector field}
In Step 2 a vector field $\mathbf{Y}$ on $\NNN$ is sought. We shall now regard the process of identifying $\YY$  through the eyes of Section \ref{sec.Turkington}.
The vector field $\YY$ is found in such a way that the trajectories on $\NNN$ corresponding to the vector field are as close as possible to the measured trajectories. The comparison can be made either on $\MMM$ (embedding trajectories on $\NNN$ into $\MMM$), or on $\NNN$ (projecting trajectories from $\MMM$ onto $\NNN$).

\subsubsection{Prediction}
Finally, Eq. (\ref{eq.Gen.y}) can be solved to obtain future trajectories on $\NNN$. The embedding then lifts the trajectories to future trajectories on $\MMM$, which is a prediction of future trajectories on $\MMM$.

\section{Illustration on learning from particle dynamics}\label{sec.ill}

Let us now illustrate the foregoing theoretical construction on a recent successful method of machine learning in dynamical systems \cite{sloshing}. The physical system under investigation is a free-surface fluid, the objective is to teach a robot how to handle it. First, we address the NI modeling of such a system. The standard modeling based on the classical fluid mechanics with the Navier-Stokes equation serving as the governing equation leads to a very complex mathematical formulation. In order to avoid the difficulties associated with numerical solutions of  partial differential equations, we choose the Lagrange formulation of fluid flows (the fluid is seen as composed of fluid particles) and then still a
simpler formulation known as
the method of Smoothed Particle Hydrodynamics (SPH), see \cite{SPH},  and the method of Smoothed Dissipative Particle Dynamics (SDPD), see \cite{Espanol-Revenga,SPH-afraid}. The data base DB presented to the robot thus consists of pseudo-experimental data. These are the fluid particle trajectories calculated as solutions to the system of ordinary differential equations serving as the governing equations in the SPH and SDPD formulations of fluid flows.

\subsection{Smoothed particle hydrodynamics}

First,  we briefly recall the SPH and SDPD methods.
Imagine a fluid motion. Instead of the usual way based on partial differential equations (e.g., Navier-Stokes equations), the fluid can be described as composed of fluid quasi-particles. Dynamics of these particles is governed by Hamilton canonical equations, which are ordinary differential equations. The particles are also equipped with their energy or entropy, which makes it possible to addresses the thermodynamic behavior, see e.g. \cite{SPH,Espanol-Revenga}.

Apart from the Hamiltonian part, the evolution equations also contain irreversible terms. These terms can be constructed by direct discretization of the continuous viscous terms (as in SPH) or by including fluctuations compatible with the continuous terms through the fluctuation-dissipation theorem (SDPD), see e.g. \cite{SPH-afraid}.

\subsection{Reduced manifold}
Let us now recall a recent successful approach to machine learning taking advantage of the GENERIC framework \cite{sloshing}.
In this approach a pseudo-experimental data of fluid motion are first acquired from an SPH simulation, having a few thousand particles, $n$ being the number of particles. The detailed manifold $\MMM$ is thus $7n$-dimensional, since each particle has its position (3), velocity (3) and energy (1). The measured states of the particles represent trajectories on $\MMM$, $G(\MMM)$.

Then three different methods searching for a suitable lower-dimen\-sional submanifold are employed, namely POD (see above), locally linear embedding (LLE) and topological data analysis (TDA). Each of the methods leads to a different manifold $\NNN$. The best performance was given by TDA, where the manifold $\NNN$ was consisting of a few particles\footnote{It is often assumed that the reduced manifold keeps the structure of a cotangent bundle, such that a reversible evolution is generated by the canonical Poisson bivector (equipped with entropy) as on the original manifold. Therefore, the reduced dynamics can be interpreted as dynamics of a lower number of (quasi-)particles, since otherwise an another Poisson bivector would have to be sought. This is not, however, strictly necessary nor a limitation of the method, see for instance \cite{GENERIC-DD} \cite{GENERIC-corrections}.} (instead of a few thousand) while still giving reasonable approximation of the pseudo-experimental data. In all the three approaches, however, the reduced manifold $\NNN$ was similar to the original high-dimensional manifold $\MMM$ in the sense that it also described pseudo-particle states (although much lower number of them). The methods provided a projection $\pi$ from $\MMM$ to $\NNN$ as well as the embedding of $\NNN$ into $\MMM$. This is the manifold recognition.

\subsection{Reduced vector field}
In Step 2 a vector field $\mathbf{Y}$ on $\NNN$ is sought. It is assumed that the vector field on $\NNN$ has the GENERIC structure
\begin{equation}\label{eq.Gen.y}
	\dot{y}^a = \dL^{ab} \dE_{y^b} + \dM^{ab}\dS_{y^b},
\end{equation}
where $\dLL$ is a Poisson bivector, $\dE$ is an energy on $\NNN$, $\dMM$ is a dissipative matrix on $\NNN$ and $\dS$ is an entropy on $\NNN$. 

%Due to the knowledge of the canonical Hamiltonian structure of dynamics on $\MMM$, the Poisson bivector $\dLL$ is chosen as canonical as well. This means that the reduced dynamics is again  dynamics of the fluid particles composing the fluid in SPH and SDPD methods except that the number of particles has been drastically reduced and the are now effective quasi-particles playing the role of avarage representatives of the original fluid particles.
Energy $\dE$ is assumed to be quadratic in $y$ so that its gradient is linear operator on $y$ (a matrix), and the same is assumed for entropy $\dS$. The dissipative matrix\footnote{corresponding to dissipation potential $\Xi=\frac{1}{2} y^*_a \dM^{ab} y^*_b$} is assumed to be piecewise constant---data are fitted by regions, not necessarily monolithically---, symmetric and positive definite. The unknown matrices $ \dE_{y^b}$, $ \dS_{y^b}$ and $\dMM$ are then fitted by least squares so that the trajectories given by integration of Eqs. (\ref{eq.Gen.y}) coincide with with projection of the measured trajectories as much as possible. Least squares can also be interpreted as a result of gradient dynamics \cite{Brockett-squares}, which means that reduction takes place in that step.

Note that Eq. (\ref{eq.Gen.y}) can be simplified to
\begin{equation}
	\dot{y}^a = \dL^{ab} \dF_{y^b} - T\dM^{ab}\dF_{y^b}
\end{equation}
for isothermal systems. Here $\dF = \dE-T\dS$ is the Helmholtz free energy. In this case only two matrices would be necessary.

\subsection{Prediction}
Finally, Eq. (\ref{eq.Gen.y}) are solved to obtain future trajectories on $\NNN$. The embedding then lifts the trajectories to future trajectories on $\MMM$, which is a prediction of future trajectories on $\MMM$, showing remarkable precision in \cite{sloshing}.

\section{Illustration on learning rigid body mechanics}\label{sec.RB}
Let us now illustrate the power of geometry when learning energy of a rigid body from observation of its instantaneous axis of rotation. Imagine a freely rotating rigid body, see e.g. \cite{landau1}. Its rotations form the group $SO(3)$, playing the role of the detailed manifold $\MMM$. Rate of rotation of the rigid body is expressed by the its angular momentum regarded from the reference frame attached to the body, $\mm$, playing the role of state variables $\xx$. By the Poisson reduction technique \cite{arnold, Marsden-Ratiu-Weinstein,Simo1988} it is possible to show that dynamics of the angular momentum is generated by non-canonical Poisson bracket
\begin{equation}\label{eq.m.PB}
\{F,G\} = -\mm \cdot \left(\frac{\partial F}{\partial \mm}\times \frac{\partial G}{\partial \mm}\right),
\end{equation}
where $F$ and $G$ are two arbitrary functions of $\mm$. This bracket implies evolution equation
\begin{equation}\label{eq.m.evo}
\dot{\mm} = \mm \times \frac{\partial E}{\partial \mm},
\end{equation}
where $E(\mm)$ is kinetic energy of the body. The energy is quadratic in $\mm$,
\begin{equation}
E = \frac{1}{2} E^{ij}m_i m_j,
\end{equation}
where the symmetric positive definite matrix $E^{ij}$ is typically diagonal, 
\begin{equation}\label{eq.E.exact}
E_{\mathrm{exact}} = \frac{1}{2}\left(\frac{m_x^2}{I_x}+\frac{m_y^2}{I_y}+\frac{m_z^2}{I_z}\right),
\end{equation}
as such suitable reference frame can always be chosen.

Any function of $\mm^2$, the Euclidean norm of $\mm$ squared, is not affected by the reversible evolution because $\{m^2, H\}=0$ for any functional $H$. The magnitude of angular momentum is thus conserved regardless the choice of energy, it is a Casimir of the Poisson bracket. Therefore, the Casimir does not affect the reversible part of the evolution and, consequently, it can not be learned from the measured trajectory.

\begin{figure}
    \begin{center}
        \includegraphics[scale=0.5]{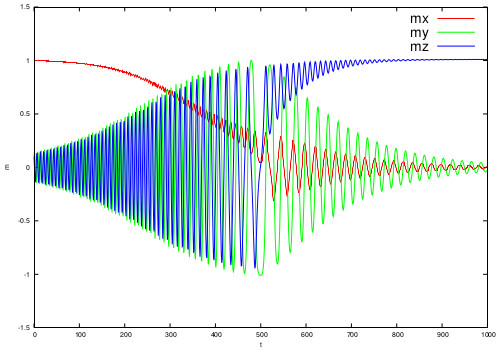}
        \caption{The rigid body starts rotating around the $x-$axis, which has the lowest moment of inertial. Due to the energetic Ehrenfest regularization in Eq. \eqref{eq.m.evo.fin} it changes its rotation to the axis with highest moment of inertia. The magnitude of angular momentum is conserved while energy is dissipated.}
        \label{fig.rot}
    \end{center}
\end{figure}

In reality, however, one typically observes not only the reversible mechanical behavior, but also the irreversible thermodynamic behavior. A rotating rigid body typically conserves its angular momentum while dissipating the kinetic energy \cite{ehrenfest-regularization}. An irreversible term referred to as the energetic Ehrenfest regularization \cite{ehrenfest-regularization}, that leads to such behavior, is added to the reversible equation \eqref{eq.m.evo}, 
\begin{equation}\label{eq.m.evo.fin}
    \dot{\mm} = \mm \times \frac{\partial E}{\partial \mm} - \frac{\tau}{2} \LL^T\cdot \frac{\partial^2 E}{\partial \mm\partial \mm} \cdot \LL \cdot \frac{\partial E}{\partial \mm},
\end{equation}
where $L^{ij} =  - m_k \epsilon^{kij}$ is the Poisson bivector generating the Poisson bracket. Equation \eqref{eq.m.evo.fin} keeps the magnitude $\mm^2$ constant while dissipating kinetic energy, $\dot{E}\leq 0$. Therefore, the rigid body tends to rotate around the axis with highest moment of inertia as in Fig. \ref{fig.rot}. By adding dissipation, the Casimirs now play a role in the dynamics, and the can be learned from the trajectory. The dissipative dynamics eventually drives the system towards an equilibrium state, where the angular momentum is aligned with the axis of highest moment of inertia. As the magnitude of the angular momentum is conserved, that state is determined uniquely and it is described by the value kinetic energy. Kinetic energy and the magnitude of angular momentum thus play the role of the lower-level state variables $\yy$. In this Section, we employ the one-level pattern recognition, where properties of dynamics on manifold $\MMM=SO(3)$ are sought while dynamics on the lower level can be obtained by projection a posteriori and does participate in the recognition process itself.

After having obtained the trajectories, we can approach the learning procedure. In the physics learning, we would like to reconstruct the formula for energy \eqref{eq.E.exact} from the trajectory of $\mm(t)$ at discrete times $t_n$. The learning is done by minimizing least squares using the Scipy method \textit{curve\_fit}. The energy is assumed to be a quadratic form of $\mm$. The Poisson bivector and time step of the numerical scheme are assumed to be known. 

\begin{figure}
    \begin{center}
        \includegraphics[scale=0.5]{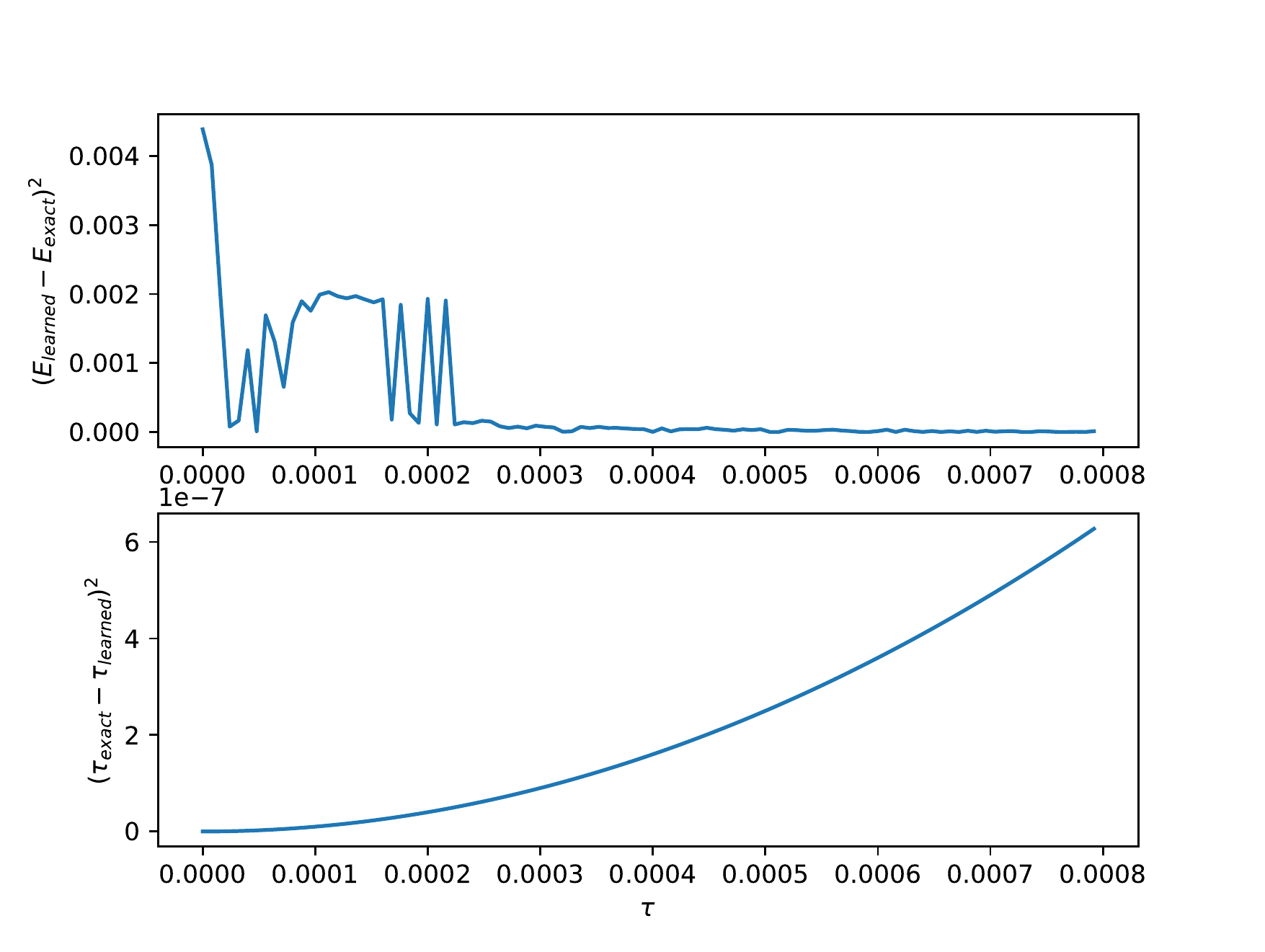}
        \caption{The absolute squared error of the fit of $\tau$ and the energy. The error of energy is calculated as the sum of squares of differences between entries of the exact and learned matrices of second derivatives of the energy. The energy is initially fitted with a significant error that eventually vanishes as the dissipation becomes significant. This means that with the dissipation we can learn the energy more accurately. The $\tau$ coefficient can be also learned with reasonable precision. We observe that the higher the exact value of the coefficient (stronger dissipation), the hither is the error between the exact and learned values.}
        \label{fig.learned.error}
    \end{center}
\end{figure}

First we tried to recognize the energy a trajectory with dissipation switched off, i.e. with $\tau=0$. As expected we could reconstruct the energy from most of the initial conditions (using only 3 subsequent values of $\mm(t)$) only up to the shift by Casimirs $\mm^2$. When using a trajectory with dissipation, $\tau > 0$, the whole energy was reconstructed. We demonstrate this on the following. We simulate the rigid body dynamics using Crank-Nicolson numerical scheme with a variable parameter $\tau$. The energetic Ehrenfest regularization is used, since it has the desired properties, conserving $\mm^2$ while dissipating kinetic energy (see \cite{ehrenfest-regularization}). The smaller $\tau$ is, the less dissipative evolution we are observing. When $\tau$ is zero, we return to the reversible Hamiltonian evolution. Therefore, it is interesting to observe what happens with the learning of energy as $\tau$ increases from zero to some small value. We let the rigid body evolve for $200$ steps while $\tau$ is in the interval $\tau \in [0, \ 8\cdot 10^{-3}]$. We then perform the learning procedure and observe error of the learned energy and also error of the learned $\tau$ compared to the exact ones (used when generating trajectories). The results can bee seen in Fig. \ref{fig.learned.error}.

In summary, knowledge of geometry makes it possible to understand and develop learning of dynamics more complex than mechanics of particles where the kinematics is not generated by Hamilton canonical equations, but by Poisson geometry with a non-canonical bracket. Such Poisson brackets involve for instance fluid mechanics, kinetic theory, electrodynamics and rigid body rotations. The latter kinematics was simulated and enhanced with terms leading to thermodynamic behavior. We find that the energy functional of a general Poisson bracket can be learned from a trajectory only up to the Casimirs of the bracket. However, when adding dissipation, the whole energy can be learned. The code is available on \cite{github-machine-learning-rigid-body}.

\section{Conclusion}
Learning is a process of getting an insight that allows to make quick predictions.
If the input of learning is a dynamical system, then the insight is an information
about important qualitative features (about a pattern) in the phase portrait (i.e.
in the collection of trajectories). One way to get such information is to reduce
the dynamical system under investigation to a simpler dynamical system whose
phase portrait is the pattern in the phase portrait corresponding to the original
dynamical system. The reduction process in which the pattern is recognized
can be interpreted as the learning process. This process can also be regarded
as a time evolution generated by a dynamics that we call a reducing dynamics
or also a learning dynamics. In the reducing time evolution the pattern in
the phase portrait of the original dynamical system emerges. The reducing
dynamics is dissipative and is driven by a potential called entropy. We can
use this terminology, since in the particular case of reductions investigated in
statistical mechanics such potentials are indeed physical entropies arising in
thermodynamics.

In the machine learning the input of learning is the phase portrait (data
base). In this paper we suggest that the approach to learning via reducing
dynamics and associated thermodynamics, that has been developed in
the context of the dynamical system theory, can also be applied and can be
useful in the machine learning. We illustrate the suggestion on the example
worked out in \cite{sloshing} and in learning energy of a rotating rigid body from observation of its angular momentum, Section \ref{sec.RB}. In particular, we show that when learning dynamics with more complex Poisson brackets than canonical, one can reconstruct the energy only up to the Casimirs of the Poisson bracket unless the dynamics is dissipative. When having both the reversible and irreversible (dissipative) terms, the whole energy can be learned from the simulated trajectory.

In summary, we display a new viewpoint of reductions that have been recently developed in non-equilibrium multiscale thermodynamics.
We argue that this new viewpoint of thermodynamics is particularly pertinent to and suitable for machine learning.
This global geometric picture linking thermodynamics and machine learning is illustrated on earlier works in this direction and on a new example using non-canonical Poisson brackets.

In the future we intend to explore  new routes opened by the connection with thermodynamics.
For instance, thermodynamics provides a close connection of entropy to fluctuations. We are suggesting that the entropy drives the learning dynamics. This means that
an appropriate analysis of fluctuations involved in the data base can serve as  a complementary tool in machine learning.

\begin{acknowledgement}
We are grateful to V{\'a}clav Klika for discussing the manuscript.

F.Ch. thanks ESI Group through its research chair at ``Arts et M\'etiers ParisTech'', whose first invited position was Prof. M. Grmela, for performing the researches here addressed. F. Ch. also knowledges Dr. Alain de Rouvray by the rich and inspiring discussions on pattern recognition as the first step towards machine learning and artificial intelligence, motivating the preset work. The support from ANR (Agence Nationale de la Recherche, France) through its grant AAPG2018 DataBEST is also gratefully acknowledged.

E.C. also acknowledges the financial support of ESI Group through the project ``Simulated Reality''. The support given by the Spanish Ministry of Economy and Competitiveness through Grant number DPI2017-85139-C2-1-R, and by the Regional Government of Aragon and the European Social Fund, research group T88, is also greatly acknowledged.

M.G. was supported by the Natural Sciences and Engineering Research Council of Canada, Grants 3100319 and 3100735.

B.M. acknowledges the support of the Spanish Ministry of Science, Innovation and Universities through grant number PRE2018-083211.

M.P. and M.{\v S} were supported by Czech Science Foundation, Project No. 20-22092S. 
M.P. was supported by Charles University Research Program No. UNCE/\-SCI/023.

\end{acknowledgement}
%
%\section*{Appendix}
%\addcontentsline{toc}{section}{Appendix}
%
%
%\input{references}

%\bibliographystyle{spphys}
%\bibliography{library} % The file containing the bibliography

\begin{thebibliography}{10}
\providecommand{\url}[1]{{#1}}
\providecommand{\urlprefix}{URL }
\expandafter\ifx\csname urlstyle\endcsname\relax
  \providecommand{\doi}[1]{DOI \discretionary{}{}{}#1}\else
  \providecommand{\doi}{DOI \discretionary{}{}{}\begingroup
  \urlstyle{rm}\Url}\fi

\bibitem{Boltzmann}
L.~Gesamtausgabe, \emph{Ludwig Boltzmann Gesamtausgabe - Collected Works}
  (1983)

\bibitem{Gorban2018_inteligence}
A.N. {Gorban}, B.~{Grechuk}, I.Y. {Tyukin}.
\newblock Augmented artificial intelligence: a conceptual framework (2018)

\bibitem{Kosambi}
D.D. Kosambi, J. Indian Math. Soc. \textbf{7}, 76 (1943)

\bibitem{Golub}
G.~Golub, C.~Van~Loan, \emph{Matrix Computations}.
\newblock Johns Hopkins Studies in the Mathematical Sciences (Johns Hopkins
  University Press, 2013).
\newblock \urlprefix\url{https://books.google.cz/books?id=X5YfsuCWpxMC}

\bibitem{LLE}
S.T. Roweis, L.K. Saul, Science \textbf{290}(5500), 2323 (2000)

\bibitem{TDA}
L.~Wasserman, Annual Review of Statistics and Its Application \textbf{5}(1),
  501 (2018)

\bibitem{Brunton}
S.~Brunton, J.~Proctor, J.~Kutz, Proceedings of the National Academy of
  Sciences  (2016).
\newblock \doi{10.1073/pnas.1517384113}

\bibitem{kaiser2018discovering}
E.~Kaiser, J.~Kutz, S.~Brunton.
\newblock Discovering conservation laws from data for control (2018)

\bibitem{kevrekidis2010equation}
Y.~Kevrekidis, G.~Samaey, Scholarpedia \textbf{5}(9), 4847 (2010)

\bibitem{E}
W.~E, Commun. Math. Stat. \textbf{5}, 1 (2017).
\newblock \doi{10.1007/s40304-017-0103-z}

\bibitem{E2018}
W.~E, J.~Han, Q.~Li, Research in the Mathematical Sciences \textbf{6}(1), 10
  (2018).
\newblock \doi{10.1007/s40687-018-0172-y}.
\newblock \urlprefix\url{https://doi.org/10.1007/s40687-018-0172-y}

\bibitem{sloshing}
B.~Moya, D.~Gonzalez, I.~Alfaro, F.~Chinesta, E.~Cueto, Computational Mechanics
  \textbf{64}(2), 511 (2019).
\newblock \doi{10.1007/s00466-019-01705-3}.
\newblock \urlprefix\url{https://doi.org/10.1007/s00466-019-01705-3}

\bibitem{Netocny2002}
C.~Maes, K.~Netočný, eprint arXiv:cond-mat/0202501  (2002)

\bibitem{go}
M.~Grmela, H.C. Öttinger, Phys. Rev. E \textbf{56}, 6620 (1997).
\newblock \doi{10.1103/PhysRevE.56.6620}.
\newblock \urlprefix\url{http://link.aps.org/doi/10.1103/PhysRevE.56.6620}

\bibitem{og}
H.C. \"Ottinger, M.~Grmela, Phys. Rev. E \textbf{56}, 6633 (1997).
\newblock \doi{10.1103/PhysRevE.56.6633}

\bibitem{hco}
H.~{\"O}ttinger, \emph{Beyond Equilibrium Thermodynamics} (Wiley, 2005)

\bibitem{pkg}
M.~Pavelka, V.~Klika, M.~Grmela, \emph{{Multiscale Thermo-Dynamics}} (De
  Gruyter, Berlin, Boston, 2018).
\newblock \doi{10.1515/9783110350951}.
\newblock
  \urlprefix\url{http://www.degruyter.com/view/books/9783110350951/9783110350951/9783110350951.xml}

\bibitem{PRE15}
M.~Pavelka, V.~Klika, M.~Grmela, Phys. Rev. E \textbf{90}(062131) (2014)

\bibitem{Onsager1930}
L.~Onsager, Phys. Rev. \textbf{37}, 405 (1931).
\newblock \doi{10.1103/PhysRev.37.405}.
\newblock \urlprefix\url{http://link.aps.org/doi/10.1103/PhysRev.37.405}

\bibitem{Onsager1931}
L.~Onsager, Phys. Rev. \textbf{38}, 2265 (1931).
\newblock \doi{10.1103/PhysRev.38.2265}

\bibitem{Casimir1945}
H.B.G. Casimir, Rev. Mod. Phys. \textbf{17}, 343 (1945).
\newblock \doi{10.1103/RevModPhys.17.343}

\bibitem{dGM}
S.R. de~Groot, P.~Mazur, \emph{Non-equilibrium Thermodynamics} (Dover
  Publications, New York, 1984)

\bibitem{RedExt}
M.~Grmela, V.~Klika, M.~Pavelka, Phys. Rev. E \textbf{92}(032111) (2015)

\bibitem{MG-CR}
M.~Grmela, Journal of Statistical Physics \textbf{166}(2), 282 (2017)

\bibitem{Miroslav-guide}
M.~Grmela, Journal of Physics Communications \textbf{2}(032001) (2018)

\bibitem{Shannon}
C.E. Shannon, Bell System Technical Journal \textbf{27}, 379 (1948)

\bibitem{Jaynes}
E.T. Jaynes, Physical Review \textbf{106}(4), 620 (1957)

\bibitem{Gorban}
A.~Gorban, I.~Karlin, \emph{Invariant Manifolds for Physical and Chemical
  Kinetics}.
\newblock Lecture Notes in Physics (Springer, 2005).
\newblock \urlprefix\url{http://books.google.cz/books?id=hjvjPmL5rPwC}

\bibitem{DynMaxEnt}
V.~Klika, M.~Pavelka, P.~V{\' a}gner, M.~Grmela, Entropy \textbf{21}(715)
  (2019).
\newblock \doi{doi:10.3390/e21070715}

\bibitem{Chapman-Cowling}
S.~Chapman, T.~Cowling, D.~Burnett, C.~Cercignani, \emph{The Mathematical
  Theory of Non-uniform Gases: An Account of the Kinetic Theory of Viscosity,
  Thermal Conduction and Diffusion in Gases}.
\newblock Cambridge Mathematical Library (Cambridge University Press, 1990).
\newblock \urlprefix\url{https://books.google.cz/books?id=Cbp5JP2OTrwC}

\bibitem{callen}
H.~Callen, \emph{Thermodynamics: an introduction to the physical theories of
  equilibrium thermostatics and irreversible thermodynamics} (Wiley, 1960).
\newblock \urlprefix\url{http://books.google.cz/books?id=mf5QAAAAMAAJ}

\bibitem{Turkington}
B.~Turkington, J Stat Phys \textbf{152}, 569 (2013)

\bibitem{JSP2020}
M.~Pavelka, V.~Klika, M.~Grmela, Journal of Statistical Physics
  \textbf{Accepted} (2020)

\bibitem{Ehrenfests}
P.~Ehrenfest, T.~Ehrenfest, \emph{The Conceptual Foundations of the Statistical
  Approach in Mechanics}.
\newblock Dover Books on Physics (Dover Publications, 1990)

\bibitem{GK-Ehrenfest}
A.N. Gorban, I.V. Karlin, H.C. \"{O}ttinger, L.L. Tatarinova, Physical Review E
  \textbf{63}(066124) (2001)

\bibitem{GK-Ehrenfest2}
I.V. Karlin, L.L. Tatarinova, A.N. Gorban, H.C. \"{O}ttinger, Physica A:
  Statistical Mechanics and its Applications \textbf{327}(3-4), 399 (2003)

\bibitem{Pavelka-LD}
M.~Pavelka, V.~Klika, M.~Grmela, Entropy \textbf{20} (2018)

\bibitem{ehrenfest-regularization}
M.~Pavelka, V.~Klika, M.~Grmela, Physica D: Nonlinear Phenomena \textbf{399},
  193  (2019).
\newblock \doi{https://doi.org/10.1016/j.physd.2019.06.006}.
\newblock
  \urlprefix\url{http://www.sciencedirect.com/science/article/pii/S0167278918305232}

\bibitem{Grmela2013-CMAT}
M.~Grmela, Computers and Mathematics with Applications \textbf{65}(10), 1457
  (2013).
\newblock \doi{http://dx.doi.org/10.1016/j.camwa.2012.11.019}.
\newblock
  \urlprefix\url{http://www.sciencedirect.com/science/article/pii/S0898122112006803}

\bibitem{Miroslav2014-Entropy}
M.~Grmela, Entropy \textbf{16}(3), 1652 (2014).
\newblock \doi{10.3390/e16031652}

\bibitem{RUR}
K.l.. {\v C}apek, \emph{R.U.R. (Rossum's Universal Robots)} (London ; New York
  :Penguin Books, 2004)

\bibitem{Chatterjee}
A.~Chatterjee, Current Science \textbf{78}(7) (2000)

\bibitem{Brockett1988}
R.~Brockett, Linear algebra and its applications \textbf{146}, 79 (1991).
\newblock \doi{10.1016/0024-3795(91)90021-N}

\bibitem{Bloch-dissipation}
A.M. Bloch, P.S. Krishnaprasad, J.E. Marsden, T.S. Ratiu, Annales de l'I.H.P.
  Analyse non lin\'eaire \textbf{11}(1), 37 (1994)

\bibitem{Absil}
A.~Absil, International Journal of Unconventional Computing \textbf{2}(4), 291
  (2006)

\bibitem{SPH}
R.~Gingold, J.~Monaghan, Mon. Not. R. Astron. Soc. \textbf{181}(3), 375 (1977)

\bibitem{Espanol-Revenga}
P.~Espa\~nol, M.~Revenga, Phys. Rev. E \textbf{67}, 026705 (2003).
\newblock \doi{10.1103/PhysRevE.67.026705}.
\newblock \urlprefix\url{https://link.aps.org/doi/10.1103/PhysRevE.67.026705}

\bibitem{SPH-afraid}
M.~Ellero, P.~Espa{\v n}ol, Appl. Math. Mech. \textbf{39}(1), 103 (2018)

\bibitem{GENERIC-DD}
D.~Gonz{\' a}lez, F.~Chinesta, E.~Cueto, Continuum Mechanics and Thermodynamics
   (2018).
\newblock \doi{10.1007/s00161-018-0677-z}

\bibitem{GENERIC-corrections}
D.~Gonz{\' a}lez, F.~Chinesta, E.~Cueto, Frontiers Materials \textbf{6}(14)
  (2019)

\bibitem{Brockett-squares}
R.~Brockett, Linear Algebra and its Applications \textbf{122-124}, 761  (1989).
\newblock Special Issue on Linear Systems and Control

\bibitem{landau1}
L.~Landau, E.~Lifshitz, \emph{Mechanics}.
\newblock Butterworth-Heinemann (Butterworth-Heinemann, 1976)

\bibitem{arnold}
V.~Arnold, Annales de l'institut Fourier \textbf{16}(1), 319 (1966)

\bibitem{Marsden-Ratiu-Weinstein}
J.~Marsden, T.~Ratiu, A.~Weinstein, Transactions of the american mathematical
  society \textbf{281}(1), 147 (1984).
\newblock \doi{10.2307/1999527}

\bibitem{Simo1988}
J.C. Simo, J.E. Marsden, P.S. Krishnaprasad, Archive for Rational Mechanics and
  Analysis \textbf{104}(2), 125 (1988).
\newblock \doi{10.1007/BF00251673}.
\newblock \urlprefix\url{https://doi.org/10.1007/BF00251673}

\bibitem{github-machine-learning-rigid-body}
M.~Pavelka, M.~{\v S}{\' i}pka.
\newblock machine-learning-rigid-body.
\newblock https://github.com/enaipi/machine-learning-rigid-body.git (2020)

\end{thebibliography}

\end{document}